\input{aipcheck}

\documentclass[
    ,final            
  ]
  {aipproc}

\layoutstyle{6x9}

\usepackage{epic,amsmath,graphicx}

\begin{document}

\title{Solution to Trouble of $D_{sJ}$ Particles
\footnote{Talk presented by T. Matsuki at XI International Conference
on Hadron Spectroscopy held at CBPF, Rio de Janeiro, Brasil,
August 21-26, 2005.}
}


\author{Takayuki Matsuki}
{
  address={Tokyo Kasei University,
1-18-1 Kaga, Itabashi, Tokyo 173, JAPAN
\footnote{E-mail: matsuki@tokyo-kasei.ac.jp}}
}

\author{Toshiyuki Morii}{
  address={Faculty of Human Development,
Kobe University,\\ Nada, Kobe 657-8501, JAPAN
\footnote{E-mail: morii@kobe-u.ac.jp}}
}

\author{Kazutaka Sudoh}{
  address={Institute of Particle and Nuclear Studies, 
High Energy Accelerator Research Organization, \\ 
1-1 Ooho, Tsukuba, Ibaraki 305-0801, JAPAN
\footnote{E-mail: kazutaka.sudoh@kek.jp}}
}

\begin{abstract}
Recent discovery of $D_{sJ}$ particles, which are considered to be a great
trouble by experimentalists as well as theorists, had already been resolved by
our potential model proposed some time ago by two of us (T.M. and T.M.), in
which the Hamiltonian and wave functions are expanded in
1/(heavy quark mass) respecting heavy quark symmetry.

Using our model, we explain how narrow states like $D_{sJ}$ can be realized,
predict properties of $0^+$ and $1^+$ states of $B$ and $B_s$ heavy mesons,
and interpret how global $SU(3)$ symmetry seems to be recovered for these
$0^+$ and $1^+$ heavy mesons.
\end{abstract}

\maketitle


\section{Introduction}
Recent discovery of narrow meson states $D_{sJ}(2317)$ by BaBar \cite{BaBar},
$D_{sJ}(2460)$ by CLEO \cite{CLEO} and the following confirmation of both
states by Belle \cite{Belle} has driven many theorists to explain these states
since the former study of these states using a semi-relativistic potential
model \cite{GIK,EGF} seems to fail to reproduce
these mass values. This triggered a series of study on spectroscopy of heavy
mesons again.
To understand these states, an interesting explanation is proposed by
Bardeen, Eichten, Hill and others \cite{BEH,BH} who
used an effective Lagrangian with heavy quark symmetry combined with chiral
symmetry of light quarks but they cannot reproduce absolute mass values
of these particles.

This problem of $D_{sJ}$ has been considered by many experimentalists as well as
theorists to be a great trouble until recently. It turns out, however, that actually we
have already predicted these masses within one percent accuracy at the first
order of purturbation in $1/m_Q$ with $m_Q$ heavy quark mass some time
ago\cite{MM} by using the same potential
model as in \cite{GIK,EGF}. Since the data adopted at the time of
publication\cite{MM} is obsolete, we are now refining calculations up to
the second order.\cite{MMS} The main difference between our treatment of the
potential
model and others is that we have taken into account negative energy states of
a heavy quark in a bound state while others do not. Namely mass is expressed
as an eigenvalue of a Hamiltonian of the heavy-light system, in which a bound
state consists of a heavy quark and a light anti-quark and negative energy
states appear in the intermediate states of a heavy quark in calculating an
energy eigenvalue, that is missed in the former study. These negative-energy
state contribution is not so small which is the main reason why people of
\cite{GIK,EGF} could not get a right answer.

Figure 1. shows how our model generates masses step by step. At first chiral
symmetry is intact with a limit of light-quark mass and scalar potential going
to zero.
At this stage no heavy quark mass correction is taken into account except for
the
lowest contribution, $m_Q$, to a heavy meson mass, i.e., heavy quark mass is
finite even at this moment. Next chiral symmetry is broken by introducing 
light-quark mass and scalar potential. There is still heavy quark symmetry
remained so that some states with partial angular momentum, $L=0,1$, between
light quark and heavy quark are still degenerate. Finally all degeneracies are
resolved by including $1/m_Q$ corrections. The similar scenario is adopted in
\cite{BEH} except for the first stage.

\begin{center}
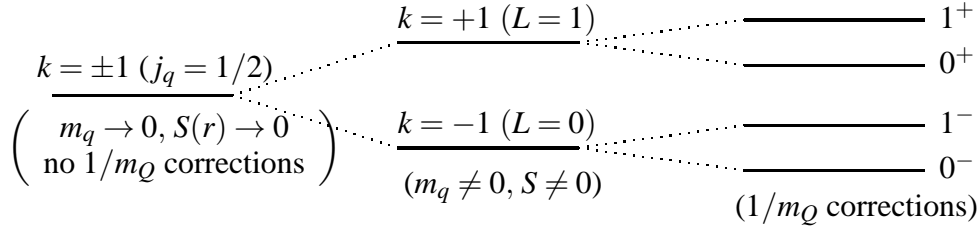
\begin{figure}[t]
\begin{picture}(370,90)
\setlength{\unitlength}{0.4mm}
 \thicklines
  \put(15,40){\line(1,0){60}}
  \put(130,57.5){\line(1,0){60}} \put(130,22.5){\line(1,0){60}}
  \put(245,65){\line(1,0){60}} \put(245,50){\line(1,0){60}}
  \put(245,30){\line(1,0){60}} \put(245,15){\line(1,0){60}}
 \thinlines
  \dottedline{3}(75,40)(130,57.5) \dottedline{3}(75,40)(130,22.5)
  \dottedline{3}(190,57.5)(245,65) \dottedline{3}(190,57.5)(245,50)
  \dottedline{3}(190,22.5)(245,30) \dottedline{3}(190,22.5)(245,15)
  \put(10,46){$k=\pm 1$ ($j_q=1/2$)}
  \put(0,20){$\left(\begin{array}{c}
              m_q\to0,\, S(r)\to0 \\  {\rm no}~ 1/m_Q~ {\rm corrections}
             \end{array}\right)$}
  \put(130,62.5){$k=+1~(L=1)$} \put(130,27.5){$k=-1~(L=0)$}
  \put(310,62.5){$1^+$} \put(310,47.5){$0^+$}
  \put(310,27.5){$1^-$} \put(310,12.5){$0^-$}
  \put(132,7.5){($m_q\neq 0,\, S\neq0$)} \put(242,0){($1/m_Q$ corrections)}
\end{picture}
\caption{Procedure how the degeneracy is resolved in our model.}
\label{mass_level}
\end{figure}
\end{center}
%

\section{Our potential model}

\subsection{$D_{sJ}$ Particles}

Our formulation using the Cornell potential is to expand Hamiltonian,
energy, and wave function in terms of $1/m_Q$ and sets coupled
euations order by order. The non-trivial differential equation is obtained
in the zeroth order, which gives orthogonal set of eigenfunctions, and
quantum mechanical perturbative corrections to energy and wave functions
in higher orders are formulated. Consistency equation is given by
\begin{eqnarray}
  && H \Psi_\ell = E^\ell \Psi_\ell, \nonumber \\
  && \left(H_{-1} + H_0 + H_1 + \cdots \right) \left(
  \Psi_{\ell 0}+\Psi_{\ell 1}+\cdots \right) =
  \left(E^\ell_0+E^\ell_1+\cdots \right) \left(
  \Psi_{\ell 0}+\Psi_{\ell 1}+\cdots \right),
\end{eqnarray}
where integers of subscripts and superscripts denote order in $1/m_Q$,
$H=H_{\rm FWT} -m_Q$, and the Foldy-Wouthuysen-Tani transformation
is operated on a heavy quark and the Hamiltonian.\cite{MM}

In Tables 1 and 2, $J^P$ stands for total spin and parity, $M_0$ lowest degenerate
mass, $c_1/M_0$ the first order correction, $M_{calc}$ calculated value of
mass, and $M_{obs}$ observed mass. In Tables 1 and 2, we give our refined
first order calculations both of $D$
and $D_s$ heavy mesons with various quantum numbers. Especially $0^+$ and
$1^+$ states
should be compared with experiments, i.e.,
$D_0^{*0}(2308)$, $D_1^{'0}(2427)$,\cite{ABE} $D_{sJ}(2317)$, and
$D_{sJ}(2460)$.\cite{BaBar,CLEO} The calculated masses,
$M_{calc}$, are within one percent of accuracy compared with the observed
masses, $M_{obs}$.
In the calculations, we have used values of parameters listed in
Table 3.

\begin{table}[t]
\caption{$D$ meson mass spectra (units are in MeV).}
\label{Dmeson}
\begin{tabular}{@{\hspace{0.5cm}}c@{\hspace{0.5cm}}r@{\hspace{1cm}}c@{\hspace{1cm}}c@{\hspace{1cm}}l@{\hspace{1cm}}c@{\hspace{1cm}}c@{\hspace{0.5cm}}}
\hline
\hline
State ($^{2s+1}L_J$) & $k$ & $J^P$ & $M_0$ & 
\multicolumn{1}{c@{\hspace{1cm}}}{$c_1 /M_0$} & 
$M_{\rm calc}$ & $M_{\rm obs}$ \\
\hline
\multicolumn{1}{@{\hspace{1cm}}l}{$^1S_0$} 
& $-1$ & $0^-$ & 1782 & 0.477 $\times 10^{-1}$ & 1867 & 1867 \\
\multicolumn{1}{@{\hspace{1cm}}l}{$^3S_1$} 
& $-1$ & $1^-$ &  & 1.272 $\times 10^{-1}$ & 2009 & 2008 \\
\multicolumn{1}{@{\hspace{1cm}}l}{$^3P_0$} 
& $1$ & $0^+$ & 2073 & 1.024 $\times 10^{-1}$ & 2285 & 2308 \\
\multicolumn{1}{@{\hspace{1cm}}l}{$"^3P_1"$} 
& $1$ & $1^+$ &  & 1.687 $\times 10^{-1}$ & 2423 & 2427 \\
\multicolumn{1}{@{\hspace{1cm}}l}{$"^1P_1"$} 
& $-2$ & $1^+$ & 2123 & 1.412 $\times 10^{-1}$ & 2423 & 2420 \\
\multicolumn{1}{@{\hspace{1cm}}l}{$^3P_2$} 
& $-2$ & $2^+$ &  & 1.616 $\times 10^{-1}$ & 2466 & 2460 \\
\multicolumn{1}{@{\hspace{1cm}}l}{$^3D_1$} 
& $2$ & $1^-$ & 2329 & 1.863 $\times 10^{-1}$ & 2763 & $-$ \\
\multicolumn{1}{@{\hspace{1cm}}l}{$"^3D_2"$} 
& $2$ & $2^-$ &  & 2.021 $\times 10^{-1}$ & 2800 & $-$ \\
\hline
\hline
\end{tabular}
\end{table}

\begin{table}[t]
\caption{$D_s$ meson mass spectra (units are in MeV).}
\label{Dsmeson}
\begin{tabular}{@{\hspace{0.5cm}}c@{\hspace{0.5cm}}r@{\hspace{1cm}}c@{\hspace{1cm}}c@{\hspace{1cm}}l@{\hspace{1cm}}c@{\hspace{1cm}}c@{\hspace{0.5cm}}}
\hline
\hline
State ($^{2s+1}L_J$) & $k$ & $J^P$ & $M_0$ & 
\multicolumn{1}{c@{\hspace{1cm}}}{$c_1 /M_0$} & 
$M_{\rm calc}$ & $M_{\rm obs}$ \\
\hline
\multicolumn{1}{@{\hspace{1cm}}l}{$^1S_0$} 
& $-1$ & $0^-$ & 1902 & 0.351 $\times 10^{-1}$ & 1969 & 1969 \\
\multicolumn{1}{@{\hspace{1cm}}l}{$^3S_1$} 
& $-1$ & $1^-$ &  & 1.101 $\times 10^{-1}$ & 2112 & 2112 \\
\multicolumn{1}{@{\hspace{1cm}}l}{$^3P_0$} 
& $1$ & $0^+$ & 2102 & 1.088 $\times 10^{-1}$ & 2330 & 2317 \\
\multicolumn{1}{@{\hspace{1cm}}l}{$"^3P_1"$} 
& $1$ & $1^+$ &  & 1.762 $\times 10^{-1}$ & 2472 & 2460 \\
\multicolumn{1}{@{\hspace{1cm}}l}{$"^1P_1"$} 
& $-2$ & $1^+$ & 2242 & 1.269 $\times 10^{-1}$ & 2527 & 2535 \\
\multicolumn{1}{@{\hspace{1cm}}l}{$^3P_2$} 
& $-2$ & $2^+$ &  & 1.462 $\times 10^{-1}$ & 2570 & 2572 \\
\multicolumn{1}{@{\hspace{1cm}}l}{$^3D_1$} 
& $2$ & $1^-$ & 2349 & 2.006 $\times 10^{-1}$ & 2821 & $-$ \\
\multicolumn{1}{@{\hspace{1cm}}l}{$"^3D_2"$} 
& $2$ & $2^-$ &  & 2.170 $\times 10^{-1}$ & 2859 & $-$ \\
\hline
\hline
\end{tabular}
\end{table}

\begin{table}[t]
\caption{Most optimal values of parameters.}
\label{parameter}
\begin{tabular*}{34pc}{c@{\extracolsep{\fill}}ccccccc}
\hline
\hline
Parameters 
& $\alpha_s$ & $a$ (GeV$^{-1}$) & $b$ (GeV) 
& $m_{u, d}$ (GeV) & $m_s$ (GeV) & $m_c$ (GeV) & $m_b$ (GeV) \\
 & 0.2620 & 1.937 & 0.07091 
& 0.00803 & 0.09270 & 1.040 & 4.513 \\
\hline
\hline
\end{tabular*}
\end{table}

\subsection{$0^+$ and $1^+$ States of $B$ and $B_s$}

\begin{table}[t]
\caption{$B$ meson mass spectra (units are in MeV).}
\label{Bmeson}
\begin{tabular}{@{\hspace{0.5cm}}c@{\hspace{0.5cm}}r@{\hspace{1cm}}c@{\hspace{1cm}}c@{\hspace{1cm}}l@{\hspace{1cm}}c@{\hspace{1cm}}c@{\hspace{0.5cm}}}
\hline
\hline
State ($^{2s+1}L_J$) & $k$ & $J^P$ & $M_0$ & 
\multicolumn{1}{c@{\hspace{1cm}}}{$c_1 /M_0$} & 
$M_{\rm calc}$ & $M_{\rm obs}$ \\
\hline
\multicolumn{1}{@{\hspace{1cm}}l}{$^1S_0$} 
& $-1$ & $0^-$ & 5255 & 0.373 $\times 10^{-2}$ & 5275 & 5279 \\
\multicolumn{1}{@{\hspace{1cm}}l}{$^3S_1$} 
& $-1$ & $1^-$ &  & 0.994 $\times 10^{-2}$ & 5307 & 5325 \\
\multicolumn{1}{@{\hspace{1cm}}l}{$^3P_0$} 
& $1$ & $0^+$ & 5546 & 0.882 $\times 10^{-2}$ & 5595 & $-$ \\
\multicolumn{1}{@{\hspace{1cm}}l}{$"^3P_1"$} 
& $1$ & $1^+$ &  & 1.453 $\times 10^{-2}$ & 5627 & $-$ \\
\multicolumn{1}{@{\hspace{1cm}}l}{$"^1P_1"$} 
& $-2$ & $1^+$ & 5596 & 1.235 $\times 10^{-2}$ & 5665 & $-$ \\
\multicolumn{1}{@{\hspace{1cm}}l}{$^3P_2$} 
& $-2$ & $2^+$ &  & 1.413 $\times 10^{-2}$ & 5675 & $-$ \\
\multicolumn{1}{@{\hspace{1cm}}l}{$^3D_1$} 
& $2$ & $1^-$ & 5802 & 1.723 $\times 10^{-2}$ & 5902 & $-$ \\
\multicolumn{1}{@{\hspace{1cm}}l}{$"^3D_2"$} 
& $2$ & $2^-$ &  & 1.870 $\times 10^{-2}$ & 5911 & $-$ \\
\hline
\hline
\end{tabular}
\end{table}

\begin{table}[t]
\caption{$B_s$ meson mass spectra (units are in MeV).}
\label{Bsmeson}
\begin{tabular}{@{\hspace{0.5cm}}c@{\hspace{0.5cm}}r@{\hspace{1cm}}c@{\hspace{1cm}}c@{\hspace{1cm}}l@{\hspace{1cm}}c@{\hspace{1cm}}c@{\hspace{0.5cm}}}
\hline
\hline
State ($^{2s+1}L_J$) & $k$ & $J^P$ & $M_0$ & 
\multicolumn{1}{c@{\hspace{1cm}}}{$c_1 /M_0$} & 
$M_{\rm calc}$ & $M_{\rm obs}$ \\
\hline
\multicolumn{1}{@{\hspace{1cm}}l}{$^1S_0$} 
& $-1$ & $0^-$ & 5375 & 0.287 $\times 10^{-2}$ & 5391 & 5369 \\
\multicolumn{1}{@{\hspace{1cm}}l}{$^3S_1$} 
& $-1$ & $1^-$ &  & 0.898 $\times 10^{-2}$ & 5424 & $-$ \\
\multicolumn{1}{@{\hspace{1cm}}l}{$^3P_0$} 
& $1$ & $0^+$ & 5575 & 0.945 $\times 10^{-2}$ & 5627 & $-$ \\
\multicolumn{1}{@{\hspace{1cm}}l}{$"^3P_1"$} 
& $1$ & $1^+$ &  & 1.531 $\times 10^{-2}$ & 5660 & $-$ \\
\multicolumn{1}{@{\hspace{1cm}}l}{$"^1P_1"$} 
& $-2$ & $1^+$ & 5715 & 1.147 $\times 10^{-2}$ & 5781 & $-$ \\
\multicolumn{1}{@{\hspace{1cm}}l}{$^3P_2$} 
& $-2$ & $2^+$ &  & 1.322 $\times 10^{-2}$ & 5791 & $-$ \\
\multicolumn{1}{@{\hspace{1cm}}l}{$^3D_1$} 
& $2$ & $1^-$ & 5822 & 1.866 $\times 10^{-2}$ & 5931 & $-$ \\
\multicolumn{1}{@{\hspace{1cm}}l}{$"^3D_2"$} 
& $2$ & $2^-$ &  & 2.018 $\times 10^{-2}$ & 5940 & $-$ \\
\hline
\hline
\end{tabular}
\end{table}

In Tables 4 and 5, we give our refined calculations both of $B$ and $B_s$
heavy mesons with various quantum numbers.
For $B$ and $B_s$ mesons, unfortunately there are only a few data. We predict
the mass of several excited states which have not yet been observed.
Among them, calculated masses of $0^+$ and $1^+$ states for $B_s$ mesons are
below $BK/B^{*}K$ thresholds the same as the case of $D_s$ mesons. 
Therefore, their decay modes to $B(0^-/1^-)+K/K^*$ are kinematically
forbidden, and the dominant modes are the pionic decay: 
\begin{eqnarray}
B_s (0^{+}) \to B_s (0^{-})+\pi , \qquad
B_s (1^{+}) \to B_s (1^{-})+\pi. 
\end{eqnarray}
The decay widths of these states are expected to be narrow as for those of $D_s$ mesons, since these decay modes violate the isospin invariance.

The similar decay modes are predicted for $B(0^+)$ and $B(1^+)$ mesons.
\begin{eqnarray}
B (0^{+}) \to B (0^{-})+\pi , \qquad
B (1^{+}) \to B (1^{-})+\pi,
\end{eqnarray}
but with broad decay widths, the same as $D(0^+)$ and $D(1^+)$ since
isospin invariance is not broken and there is no threshold.

These higher states of $B$ and $B_s$ mesons might be observed in Tevatron/LHC experiments in near future by analyzing above decay modes. 
We hope that our framework is confirmed in the forthcoming experiments.

\subsection{Recovery of Global $SU(3)$ for $0^+$ and $1^+$ Heavy Mesons}

When one carefully looks at Tables 1 and 2, one notices that it looks that
the global $SU(3)$ flavor symmetry is recovered when $J^P=0^+$
and $1^+$ for $D_{(s)}$ mesons compared with $0^-$ and $1^-$.
The magnitude of $SU(3)$ symmetry breakdown for $D_{(s)}(0^-)$ and
$D_{(s)}(1^-)$ mesons is given by (units in MeV),
\begin{equation}
  M(D(c\bar u)) \approx M(D(c\bar d)) \approx M(D(c\bar s)) - 100,
\end{equation}
while the symmetry seems to be recovered for $D(0^+)$ and $D(1^+)$ mesons,
\begin{equation}
  M(D(c\bar u)) \approx M(D(c\bar d)) \approx M(D(c\bar s)) - (9\sim 33).
\end{equation}
For instance, mass difference among members of a multiplet
$(D(0^-), D_s(0^-))$ seems to be larger compared with that among those
of $(D(0^+), D_s(0^+))$.
These equations are both for observed and calculated masses.\cite{MM, MMS}
What causes this recovery of $SU(3)$ symmentry? Is this just an accidental or
is there a deep meaning for this?
We study this phenomenon in details\cite{MMS2} and have found that this is a
general phenomenon occuring for heavy mesons, $(Q\bar q)$, as well as heavy
baryons like $(QQq)$.

This is explained by drawing a figure of mass gap between degenerate
states, $M_0(0^-)=M_0(1^-)~(L=0)$ and $M_0(0^+)=M_0(1^+)~(L=1)$.
These degenerate masses are written as $M_0=m_Q+C_0(m_q)$, i.e., a function
of light quark mass $m_q$. Hence defining
\begin{eqnarray*}
  \Delta M(m_u) = M_{D 0}(0^+)-M_{D0}(0^-), \quad
  \Delta M(m_s) = M_{D_s 0}(0^+)-M_{D_s 0}(0^-),
\end{eqnarray*}
these mass gaps have the same values for $D$ and $B$ mesons. This mass gap
function has {\it monotonous decreasing tendency} so that
$\Delta M(m_u) > \Delta M(m_s)$ for $m_u < m_s$.
\begin{eqnarray}
  \Delta M(m_s) - \Delta M(m_u) = -91 {\rm ~MeV},
\end{eqnarray}
which almost cancels mass difference between $D(c\bar u)$ ($D(c\bar d)$) and
$D(c\bar s)$ given by Eq.(4). Equation (6) is the origin of recovery of
$SU(3)$ invariance, which does not depend on whether the light-heavy system
is $Q\bar q$ or $QQq$ when $q=u,d,s$. That is, this is a general phenomenon
not peculiar to $D(0^\pm)$ and $D(1^\pm)$. Relation between this phenomenon
and constituent quark mass will be discussed elsewhere.\cite{MMS2}



\def\Journal#1#2#3#4{{#1} {\bf #2}, #3 (#4)}
\def\NIM{Nucl. Instrum. Methods}
\def\NIMA{Nucl. Instrum. Methods A}
\def\NPB{Nucl. Phys. B}
\def\PLB{Phys. Lett. B}
\def\PRL{Phys. Rev. Lett.}
\def\PRD{Phys. Rev. D}
\def\ZPC{Z. Phys. C}
\def\EPJ{Eur. Phys. J. C}
\def\PR{Phys. Rept.}
\def\IJM{Int. J. Mod. Phys. A}



\begin{thebibliography}{00}
\newcommand{\etal}{{\it et al.}}
\bibitem{BaBar} BaBar Collaboration, B. Aubert \etal, 
                \Journal{\PRL}{90}{242001}{2003}.
\bibitem{CLEO} CLEO Collaboration, D. Besson \etal, 
               \Journal{\PRD}{68}{032002}{2003}.
\bibitem{Belle} Belle Collaboration, Y. Mikami \etal, 
               \Journal{\PRL}{92}{012002}{2004}.
\bibitem{GIK} S. Godfrey and N. Isgur, 
             \Journal{\PRD}{32}{189}{1985};
              S. Godfrey and R. Kokoski, 
             \Journal{\PRD}{43}{1679}{1991}.
\bibitem{EGF} D. Ebert, V.O. Galkin, and R.N. Faustov,
             \Journal{\PRD}{57}{5663}{1998}.
\bibitem{BEH} W. A. Bardeen, E. J. Eichten, and C. T. Hill,
             \Journal{\PRD}{68}{054024}{2003}.
\bibitem{BH} W. A. Bardeen and C. T. Hill, 
             \Journal{\PRD}{49}{409}{1994}; 
             M. A. Nowak, M. Rho, and I. Zahed, 
             {\it ibid.} {\bf 48}, 4370 (1993); 
             A. Deandrea, N. Di Barolomeo, R. Gatto, G. Nrdulli, 
             and A. D. Plosa, 
             {\it ibid.} {\bf 58}, 034004 (1998); 
             A. Hiorth and J. O. Eeg, 
             {\it ibid.} {\bf 66}, 074001 (2002).
\bibitem{MM}
    T. Matsuki and T. Morii, \Journal{PRD}{56}{5646}{1997};
    T. Matsuki, K. Mawatari, T. Morii, and K. Sudoh,
    \Journal{\PLB}{606}{329}{2005}.
\bibitem{MMS}
    T. Matsuki, T. Morii, and K. Sudoh, work in progress.
\bibitem{ABE} Belle Collaboration, K. Abe \etal,
   \Journal{\PRD}{69}{1120022}{2004};
   Talk given by K. Abe at the {\sl International Workshop PENTAQUARK04} held
   at Spring-8 in July 19-23, 2004.
\bibitem{MMS2}
    T. Matsuki, T. Morii, and K. Sudoh, work in progress.
\end{thebibliography}
\end{document}